\documentclass{ptapap}

\author{J. Daszy\'nska-Daszkiewicz}[IAUWr]
\author{P. Walczak}[IAUWr]
\author{A. Pamyatnykh}[CAMK]
\author{M. Jerzykiewicz}[IAUWr]
\author{A. Pigulski}[IAUWr]
\author{BEST}[BEST]

\affil[IAUWr]{Instytut Astronomiczny, Uniwersytet Wroc{\l}awski, Kopernika 11, 51-622 Wroc{\l}aw, Poland}
\affil[CAMK]{Nicolaus Copernicus Astronomical Center, ul. Bartycka 18, 00--716 Warszawa, Poland}
\affil[BEST]{Bright Target Explorer (BRITE) Executive Science Team}

\title{The solitary g-mode frequencies in early B-type stars}

\begin{document}

\maketitle

\begin{abstract}
We present possible explanations of pulsations in early B-type main sequence stars
which arise purely from the excitation of gravity modes.
There are three stars with this type of oscillations detected from the BRITE light curves: $\kappa$ Cen, a Car, $\kappa$ Vel.
We show that by changing metallicity or the opacity profile it is possible in some models to dump pressure modes
keeping gravity modes unstable. Other possible scenario involves pulsations of a lower mass companion.
\end{abstract}

\section{Introduction}

For many years, it seemed that B-type main sequence (MS) pulsators can be divided into two groups:
$\beta$ Cephei stars with low-order p/g-mode pulsations (e.g. Moskalik \& Dziembowski 1992, Stankov \& Handler 2005) and Slowly Pulsating B-type stars (SPB)
with high-order g-mode pulsations (Waelkens 1991, Dziembowski, Moskalik \& Pamyatnykh 1993)).
This simple dichotomy was spoiled by the continuous increase in accuracy of the time-series photometry
obtained from multi-site campaigns and, mainly, from space missions (CoRoT, Kepler, BRITE).
Thereby, many pulsators with so far unknown oscillation spectra were identified.
The first example are hybrid pulsators with low order p/g modes and high-order g modes excited simultaneously.
Then, even more unexpected by theoretical models oscillation spectra were obtained from the CoRoT and Kepler space data:
late B-type stars with only high frequencies and early B-type stars with only low frequencies (Degroote et al. 2009, Balona et al. 2011, 2015).
All these three groups pose a challenge for theoretical modelling of stellar pulsations.

Recently, early B-type stars with solitary g mode frequencies were also identified from the BRITE photometry.
Here, we present interpretation of this case, exemplified by the three stars observed
by the BRITE satellites: $\kappa$ Cen, a Car, $\kappa$ Vel.

In Sect.\,2, we give basic information about the stars. In Sect.\,3 we search for pulsational models
of early B-type stars which can account for the observed frequencies in these three stars.
Sect.\,3 contains other possible explanations. We end with Summary and prospects for further works.

\section{The objects of interest}

According to the spectral classification, all three stars are main sequence objects of early B spectral type.
Their basic parameters are given in Table\,1. The subsequent columns include: spectral type, the brightness in the Johnson $V$ passband,
effective temperature and luminosity derived from the mean photometric indices and Hipparcos parallaxes, and the value of  $V_{\rm rot}\sin i$ from the literature.
Moreover, the last two columns contain the information about the number of frequency peaks identified in the BRITE light curves
and the range of these frequencies.

\begin{table*}
\begin{center}
\caption{Basic data for the studied stars.}
\begin{tabular}{|r|c|c|c|c|c|c|c|c|}
\hline
  Star    &   SpT  & $m_V$ & $T_{\rm eff}$ & $\log L/L_\odot$ & $V_{\rm rot}\sin i$ &  N & freq. range   \\
          &        & [mag] &      [K]      &                  &    [km/s] &    & [d$^{-1}$] \\
\hline
\hline
$\kappa$ Cen &  B2IV  & 3.1  &   $19800\pm900$   &    $3.4\pm0.2$   &     $11\pm7^1$     & 12 & (0.33, 1.37) \\
\hline
       a Car & B2IV-V & 3.4  &   $18000\pm600$   &    $3.5\pm0.1$   &     $40\pm7^2$     &  5 & (0.37, 1.71) \\
\hline
$\kappa$ Vel & B2IV-V & 2.5  &   $19400\pm600$   &    $4.2\pm0.1$   &      $\sim 49^3$   &  8 &  (0.17, 0.66) \\
\hline
\end{tabular}
{\small (1) Hubrig et al. (2009); (2) Bernacca \& Perinotto (1970); (3) Leone \& Lanzafame (1997)}
\end{center}
\end{table*}

None of these stars is single. $\kappa$ Cen is most probably a physical triple system (Kouwenhoven et al. 2007).
Visually, the bright star, $\kappa$ Cen A is paired with the weak component $\kappa$ Cen B (11th magnitude)
which is a dwarf of the spectral class K2.
Then, $\kappa$ Cen A is a spectroscopic binary and splits into Aa (the B2IV-type star) and Ab (probably a class A0V).
The primary Aa was a candidate $\beta$ Cephei variable that showed line-profile variations (Schrijvers et al. 2002).

The next object, a Car, is a single-lined spectroscopic binary (SB1)with the orbital period $P_{\rm orb}=6.74483(9)$ d,
the radial-velocity amplitude $K=18.2(2.0)$ km/s and the eccentricity of the orbit $e=0.34(14)$ (Monet 1980).
$\kappa$ Vel is also a binary of the SB1 type. The orbital period $P_{\rm orb}=116.65$ d, the radial velocity amplitude $K=46.5$ km/s
and eccentricity $e=0.19$ were derived by Curtis (1907).
\begin{figure}[h]
\centering
\begin{minipage}{0.48\textwidth}
   \includegraphics[clip,width=\textwidth]{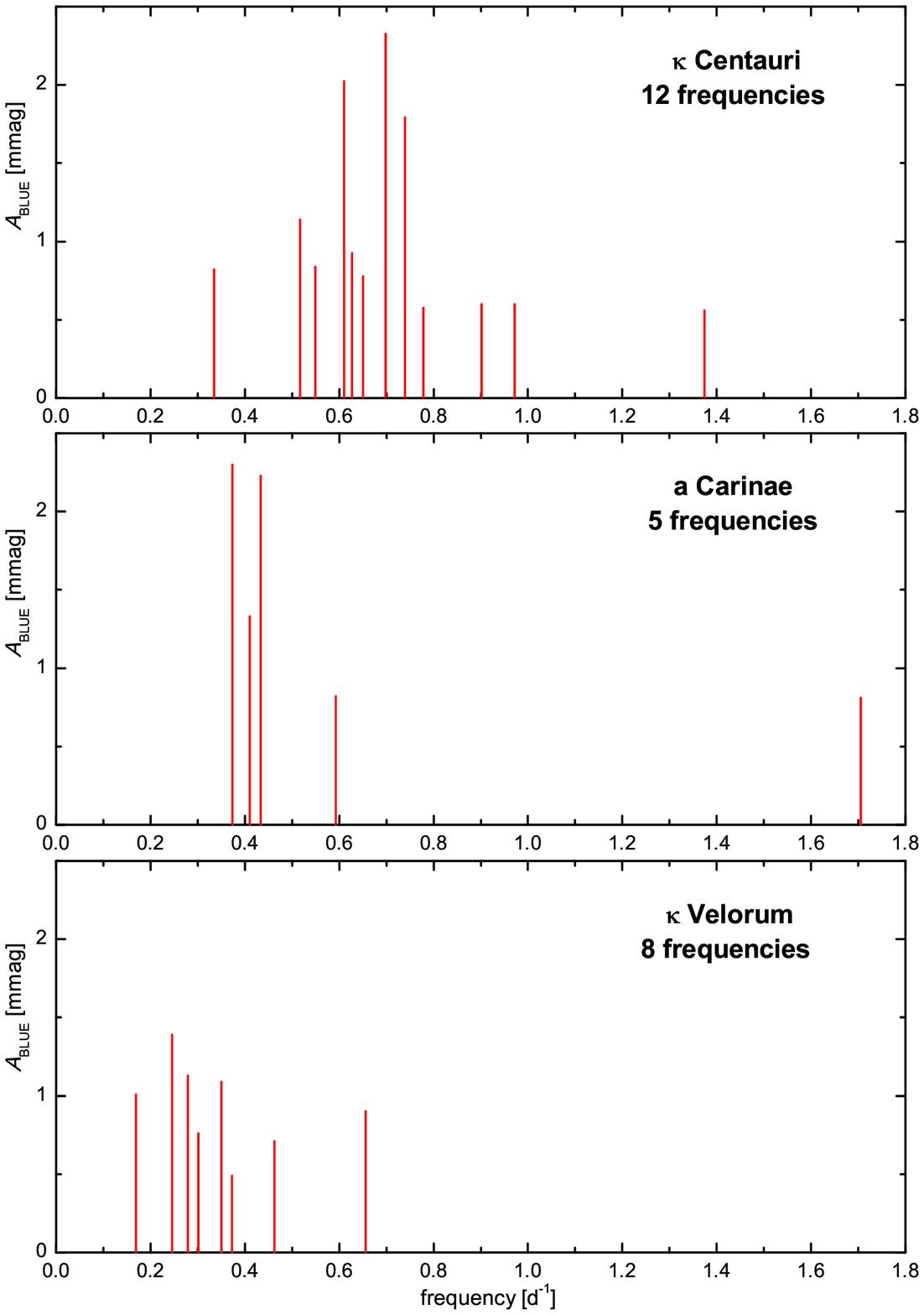}
   \caption{The oscillation spectra of $\kappa$ Cen, a Car and $\kappa$ Vel obtained from the BRITE photometry.}
   \label{fig:osc_freq}
\end{minipage}
 \quad
  \begin{minipage}{0.48\textwidth}
    \includegraphics[clip,width=\textwidth]{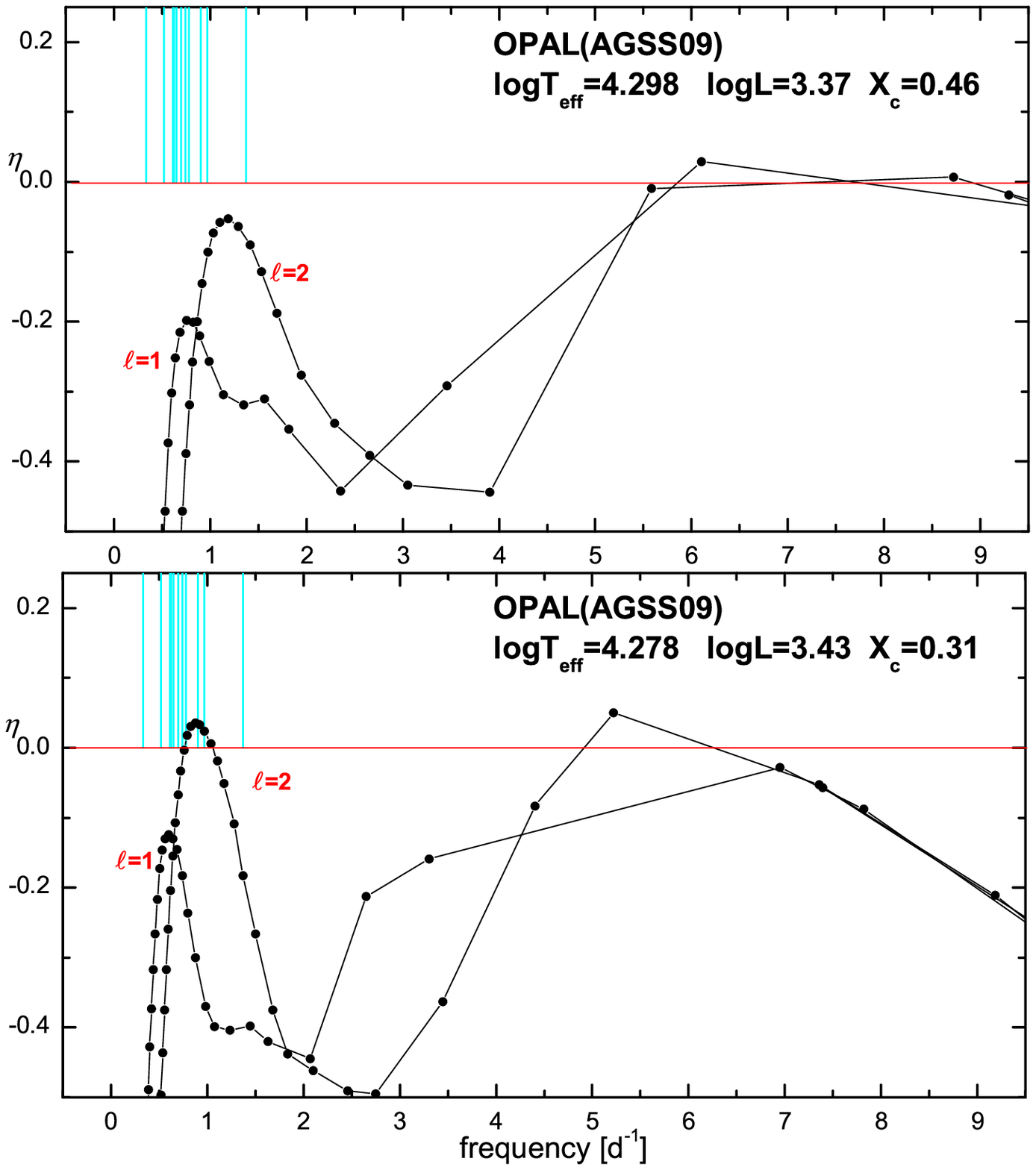}
    \caption{The normalized instability parameter, $\eta$, as a function of the mode frequency for two OPAL models with a mass of $M=7~M_\odot$ suitable for $\kappa$ Cen.
     The model in the bottom panel is more evolved. Pulsational modes with the degree $\ell=1,2$ are shown. The observed frequency peaks are plotted with the vertical lines.}
    \label{fig:fig2}
  \end{minipage}
\end{figure}

BRITE photometry of all stars was corrected for instrumental effects which included outlier rejection and decorrelations with centroid positions and CCD temperature. Finally, offsets between individual setups were accounted for and such data were subject of the subsequent time-series analysis.
The oscillation spectra of the three stars derived from the combined blue and red BRITE magnitudes are shown in Fig.\,1.
In total, twelve, five and eight terms with amplitudes above the detection threshold were found in the photometry of Kappa Cen, a Car, and Kappa Vel, respectively.
The frequency $\nu_4=0.59294$ d$^{-1}$ of a Car is equal to $4\nu_{\rm orb}$ with an accuracy better than $1\sigma$.
Thus, $\nu_4$ could be a tidally excited mode given rather high eccentricity $e=0.34$ and not very long orbital period $P_{\rm orb}\approx 6.7$ d.

\section{Pulsational models}

\subsection{Models with standard opacities}
In the first step, we computed pulsational models with standard opacity data
commonly used in pulsational computations, i.e., OPAL and OP.
To this end, we used the Warsaw-New Jersey evolutionary code and nonadiabatic pulsational code of Dziembowski (1977).

In Fig.\, 2, we plot the normalized instability parameter, $\eta$, as a function of the mode frequency
for two main sequence models with $M=7~M_\odot$ computed with the OPAL data for the initial hydrogen abundance $X_0=0.7$,
metallicity $Z=0.015$ and the solar element mixture of Asplund et al. (2009). The positive values of $\eta$ are for unstable modes.
The two models differ in the effective temperature and have stellar parameters suitable for $\kappa$ Cen.
The frequency peaks of $\kappa$ Cen are marked as the vertical lines.

This figure recalls the important features of gravity modes in massive main sequence models.
Firstly, the instability of gravity modes increases with evolution on the main sequence
and the maximum of $\eta(\nu)$ for a given degree $\ell$ shifts to lower frequencies.
Secondly, the instability increases with $\ell$ and is shifted towards
the higher frequencies with $\ell$.
One can also see that the oscillation spectrum of $\kappa$ Cen cannot be explained by such models
because the g modes, especially with the lowest frequencies, are quite far from the instability.
Moreover, the p modes, which are not observed, remain unstable.

Computations with the OP opacities yield higher instability of g modes,
because the metal bump is located deeper than in the OPAL opacities, but still p modes are unstable.
To stabilize the p modes, we lowered metallicity to $Z=0.012$. This does not affect so much the g modes
because their instability is more resistant to the metallicity reduction (e.g., Pamyatnykh 1999).
The result is shown in Fig.\,3, where we plot $\eta$ of the OP model computed with the metallicity $Z=0.012$.
\begin{figure}[h]
  \centering
\begin{minipage}{0.48\textwidth}
    \includegraphics[width=\textwidth]{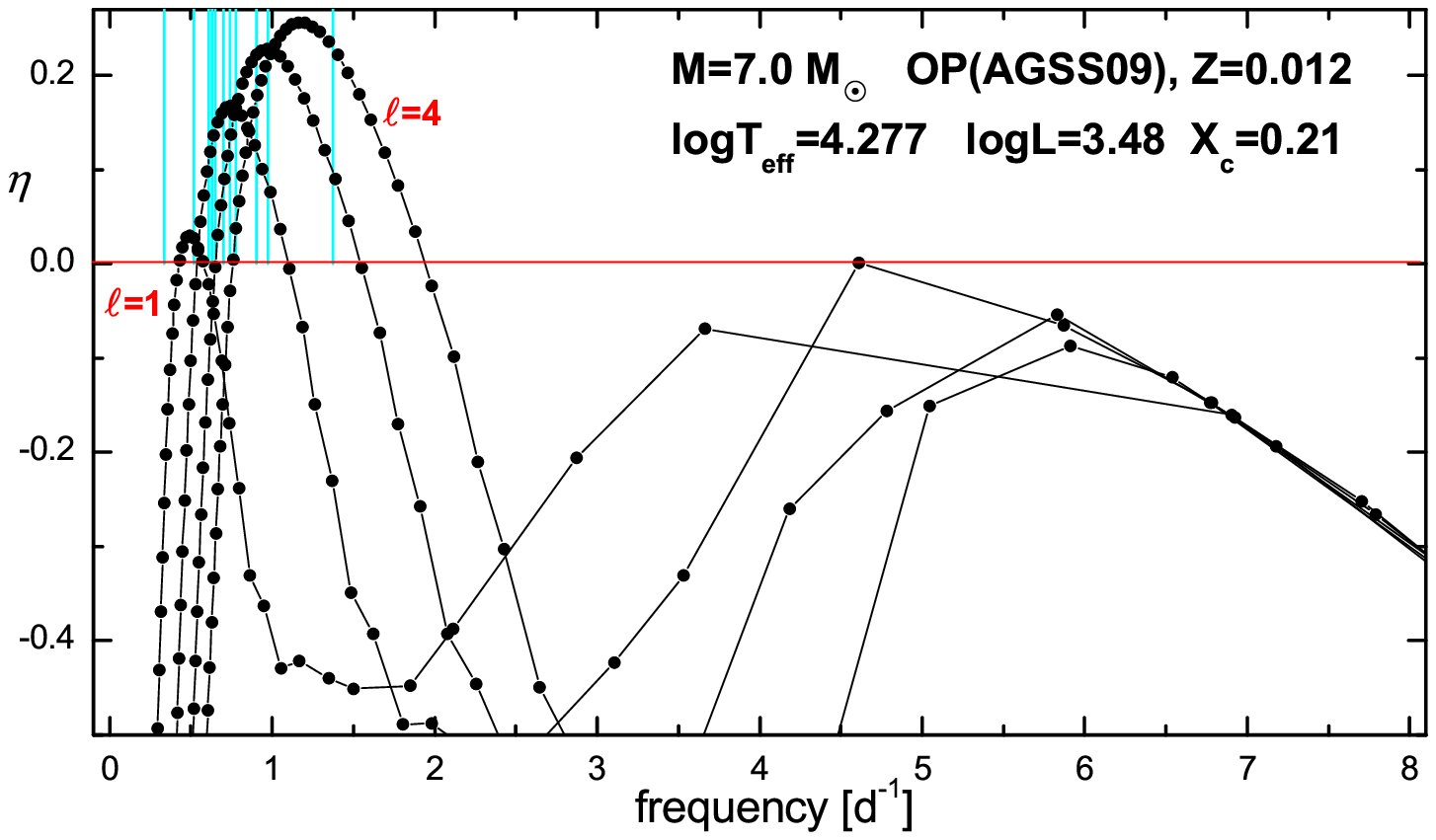}
    \caption{The same as in Fig.\,2 but for the model computed with the OP data and metallicity $Z=0.012$. Pulsational modes with the degree up to $\ell=4$ are shown.}
    \label{fig:fig3}
  \end{minipage}
  \quad
  \begin{minipage}{0.48\textwidth}
    \includegraphics[width=\textwidth]{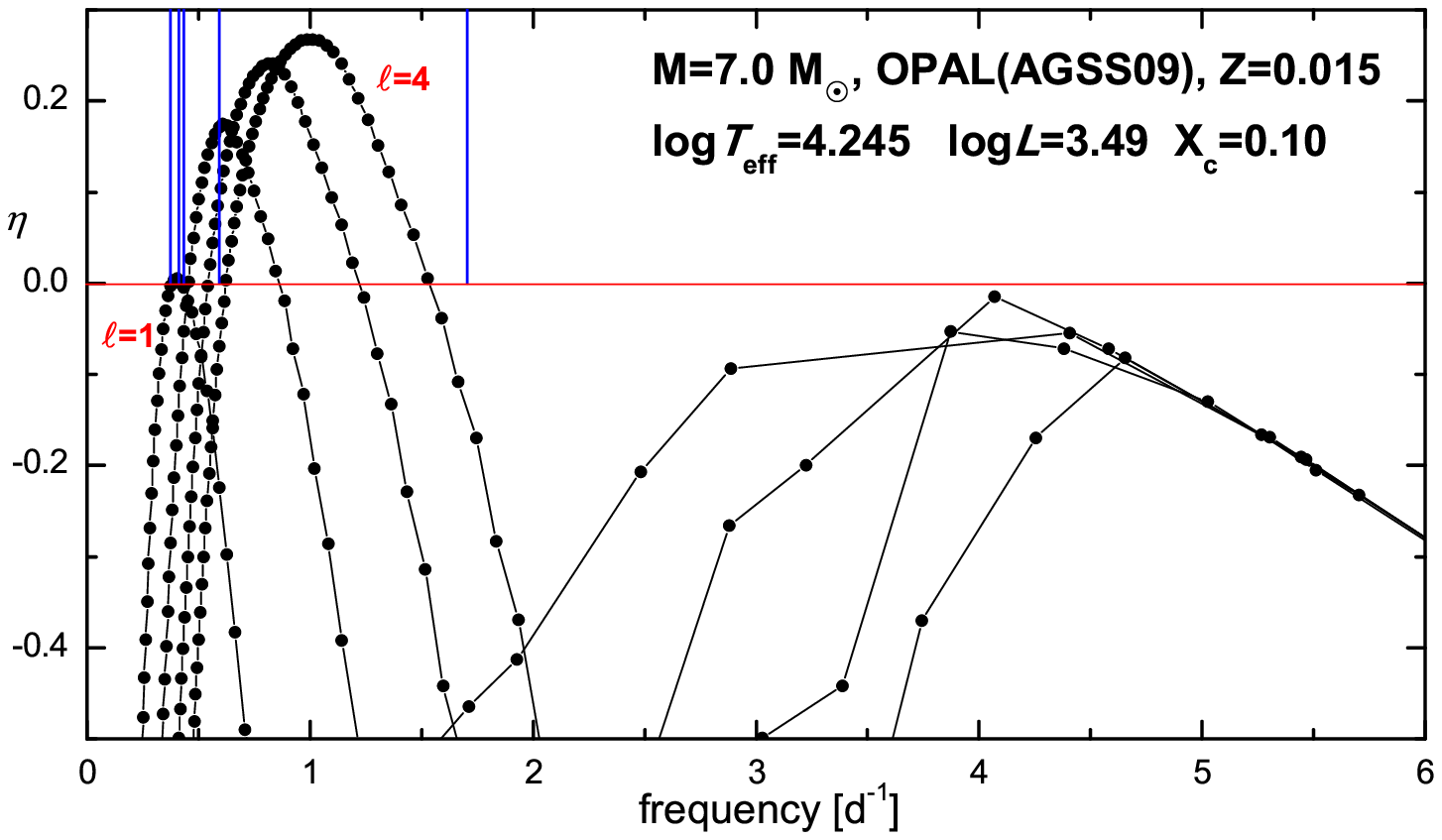}
    \caption{A comparison of the oscillation frequencies of a Car with the pulsational model computed with the OPAL data and metallicity $Z=0.015$. Modes with the degree up to $\ell=4$ are
    shown. The model has a mass of $M=7~M_\odot$ and is more evolved than the models of $\kappa$ Cen in Fig.\,2.}
    \label{fig:fig4}
  \end{minipage}
\end{figure}

Similarly to $\kappa$ Cen, a Car has a mass of about 7 $M_\odot$ but is slightly more evolved.
As mentioned above, the instability of gravity modes increases with evolution on the MS. It appeared that even with the OPAL opacities
it is possible to find pulsational models which can account for the observed frequency range of a Car.
Fig.\,4 shows a comparison of the a Car oscillation spectrum with $\eta(\nu)$ of a $M=7~M_\odot$ model computed with the OPAL opacities and metallicity $Z=0.015$.
The highest frequency mode can be explained by rotational splitting.

$\kappa$ Vel is quite a problematic object as the observed parameters do not locate it on the main sequence.
One would need much higher rotation, or unreliable overshooting and/or metallicty to put this star on the MS.
The estimated mass of $\kappa$ Vel is about 10 $M_\odot$. If the star were a post-MS object
then the observed oscillations could not be explained as well because pulsational modes with such low frequencies
would be stable in the post-MS models in the considered range of masses.

\subsection{Models with modified opacities}

Because of the still existing uncertainties in the opacity data, we computed models with a modified profile
of the mean opacity at the depths that can affect the pulsational driving of early B-type star models.
Our modification consists in an artificial changing of the opacity profile at certain depths inside the star, i.e., at certain $\log T$.

The pulsations of B-type main sequence stars are excited by the $\kappa$-mechanism operating in the Z-bump around $\log T=5.3$.
The most important contributions to the Z-bump opacity come from Fe, Ni, Mn and Cr.
The maximum opacity of these elements occurs at various depths expressed usually by $\log T$, e.g.,
the maximum contribution of Ni to the Z-bump is at $\log T=5.46$.
In Figs.\,5 and 6, we plot the instability parameter $\eta$ for a model computed with the modified opacities
which were reduced by 20\% at $\log T=5.3$ and increased by 50\% at $\log T=5.46$.
\begin{figure}[h]
  \centering
\begin{minipage}{0.48\textwidth}
    \includegraphics[width=\textwidth]{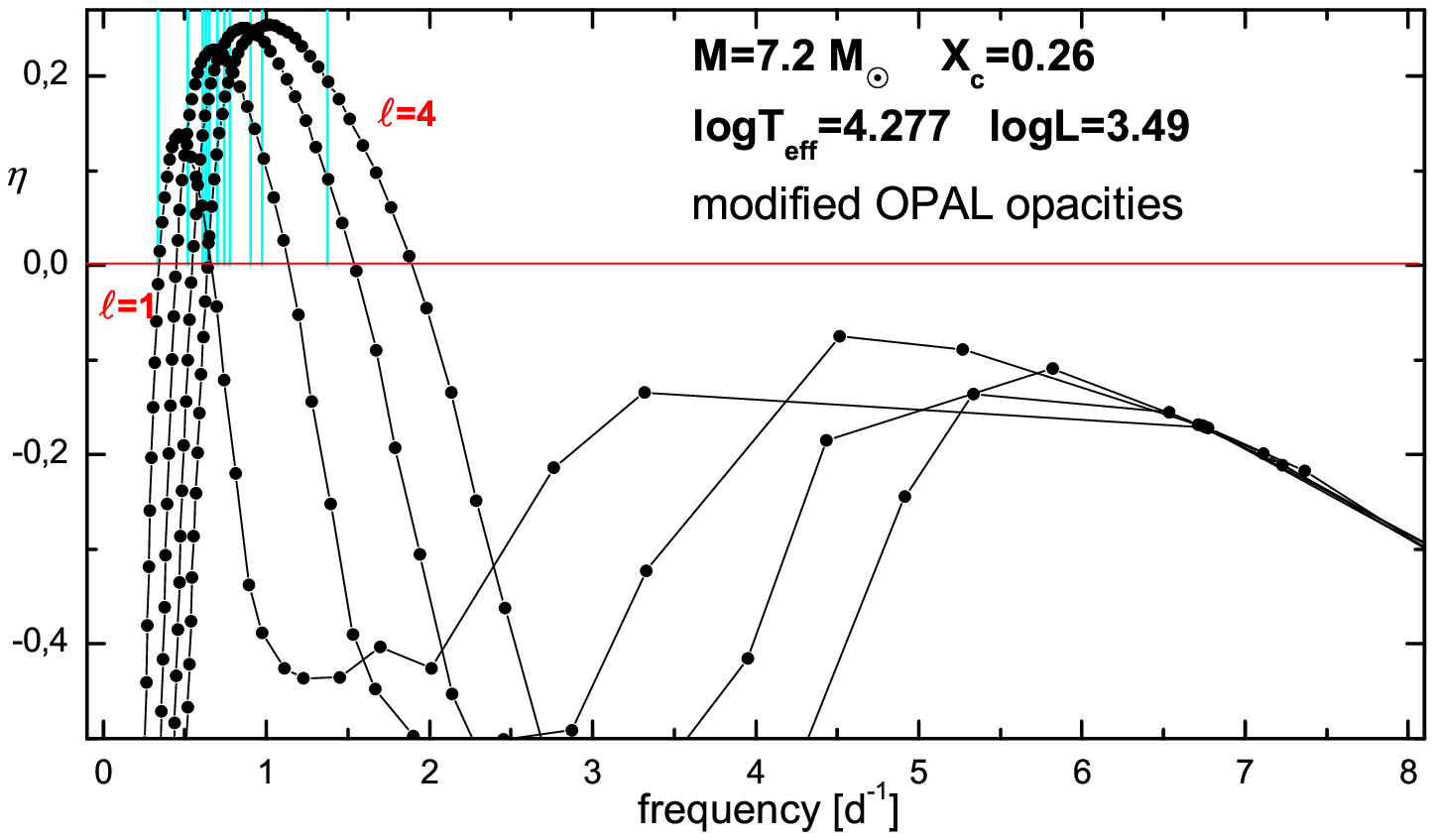}
    \caption{The parameter $\eta$ as a function of the mode frequency for a $M=7.2~M_\odot$ model computed with the modified OPAL opacities and metallicity $Z=0.015$. The opacities were increased by 50\% at $\log T=5.46$ and reduced by 20\% at $\log T=5.3$.}
    \label{fig:fig5}
  \end{minipage}
  \quad
  \begin{minipage}{0.48\textwidth}
    \includegraphics[width=\textwidth]{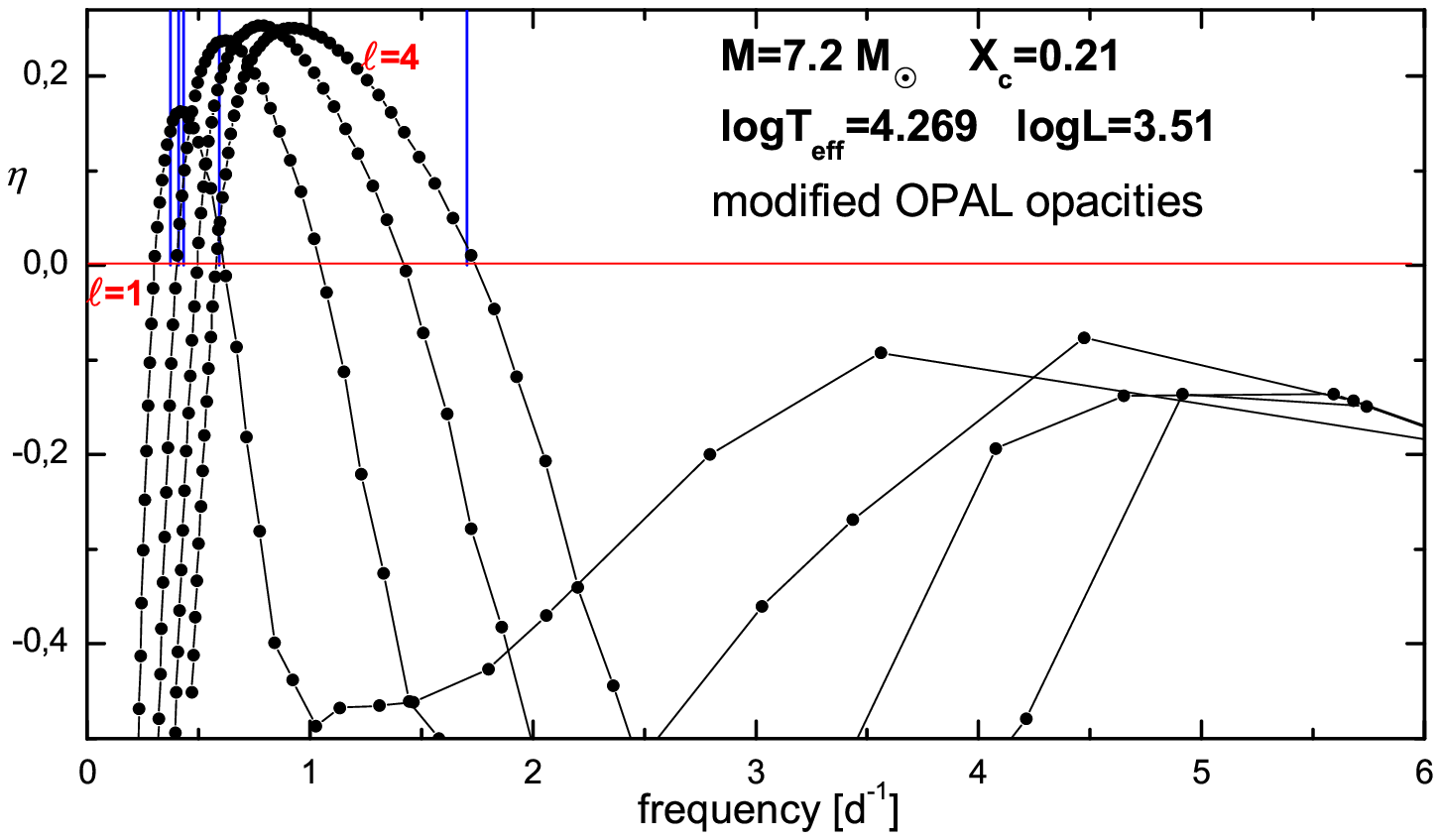}
    \caption{The same as in Fig.\,5 but for a more evolved model suitable for a Car.}
    \label{fig:fig6}
  \end{minipage}
\end{figure}

Decreasing the opacity at $\log T =5.3$ is needed to stabilize higher frequency modes
whereas to maintain and increase the instability of low frequency modes one needs to increase
the opacity at the deeper layer with temperature $\log T=5.46$.

\section{Other possible scenarios}
There is one more possible scenario.
$\kappa$ Cen is a triple system and if the Ab component, which is possibly the A0 main sequence star, has a mass of about 3.0 $M_\odot$
then the g-mode pulsations may come from this component (e.g., Walczak et al. 2015, Szewczuk \& Daszynska-Daszkiewicz 2016).
Similarly, in the case of a Car pulsation may originate from the companion.
In the case of $\kappa$ Vel, the pulsations of the lower mass companion seem to be the most credible explanation.
This component could be invisible because: 1) it rotates very fast or 2) it is too faint.
If the component is fainter by ~2 mag then there is even no need for fast rotation.
From the spectroscopic orbit we get a lower limit for the mass of the unseen component.
Assuming the orbital inclination angle $i=90^\circ$, we obtain 6.2 $M_\odot$ if the mass of the primary is $M=8~M_\odot$
and 7.0 $M_\odot$ if the mass of the primary is $M=10~M_\odot$. Solitary g modes can be easily
explained in models with masses $M\in (6 - 7) ~M_\odot$.

\section{Summary and further works}

We investigated models of early B-type stars in order to explain the oscillation spectra derived from the BRITE data
for the three stars: $\kappa$ Cen , a Car and $\kappa$ Vel. All of them exhibit only low frequency peaks although,
according to the spectral classification, they are in the main sequence stage, so that the p mode frequencies would be expected.

Our calculations may help explaining the solitary low frequency modes in the MS stellar models we consider.
The well-known facts are: 1) g modes are more unstable in more evolved models; 2) instability of g modes increases with the mode degree, $\ell$,
and is shifted to higher frequencies with $\ell$; 3) the instability of g-modes in the OP models is higher than in the OPAL models.
The instability of g modes is more resistant to the metallicity reduction comparing to the p mode instability,
and in the case of a Car this effect is sufficient to account for the oscillation frequencies.
Moreover, we examined the effects of increasing the opacities in stellar models representative for the studied stars.
We found that by decreasing the opacity at $\log T=5.3$ and increasing at $\log T=5.46$
it is possible to ``kill'' the instability of the p modes keeping the g modes unstable.
The models with such modification of the mean opacity profile explain the observed frequencies of $\kappa$ Cen.
We confirmed the Salmon's et al. (2012) result that increasing opacity at $\log T=5.46$ is crucial for g mode instability.

Because the stars we studied are spectroscopic binaries, we cannot exclude that the pulsations come from the unseen components.
In the case of $\kappa$  Vel, the pulsations of the lower mass companion seem to be the most plausible explanation of the observed frequencies.

\acknowledgements{The work was financially supported by the Polish NCN grant
\newline 2015/17/B/ST9/02082.
The paper is based on data collected by the BRITE Constellation satellite mission, designed, built, launched, operated and supported by the Austrian Research Promotion Agency (FFG), the University of Vienna, the Technical University of Graz, the Canadian Space Agency (CSA), the University of Toronto Institute for Aerospace Studies (UTIAS), the Foundation for Polish Science \& Technology (FNiTP MNiSW), and National Science Centre (NCN).}


\small
\begin{thebibliography}{}
  \bibitem{} Asplund, M., Grevesse, N., Sauval, A. J., Scott, P., 2009, ARA\&A 47, 481
  \bibitem{} Balona, L. A., Pigulski, A., Cat, P. De, Handler, G., et al. 2011, MNRAS, 413, 2403
  \bibitem{} Balona, L. A., Baran, A. S., Daszy\'nska-Daszkiewicz, J., De Cat, P., 2015, MNRAS, 451, 1445
  \bibitem{} Bernacca, P. L., Perinotto, M., 1970, Contrib.Os.Astrofis.Asiago, Univ.Padova, No.239
  \bibitem{} Curtis, H. D., 1907, PASP, 19, 259
  \bibitem{} Degroote, P., Aerts, C., Ollivier, M. et al., 2009, A\&A, 506, 47
  \bibitem{} Dziembowski, W. A., 1977, AcA, 27, 95
  \bibitem{} Dziembowski, W. A., Moskalik, P., Pamyatnykh, A. A., 1993, MNRAS, 265, 588
  \bibitem{} Hubrig, S., Briquet, M., De Cat, P., 2009, AN, 330, 317
  \bibitem{} Kouwenhoven, M. B. N., Brown, A. G. A., Portegies Zwart, S. F., Kaper, L., 2007, A\&A, 474, 77
  \bibitem{} Leone, F., Lanzafame, A. C., 1997, A\&A, 320, 893
  \bibitem{} Monet, D. G., 1980, ApJ, 237, 513
  \bibitem{} Moskalik, P., Dziembowski, W. A., 1992, A\&A, 256, L5
  \bibitem{} Pamyatnykh, A. A., 1999, Acta Astron., 49, 119
  \bibitem{} Salmon, S., Montalban, J., Morel, T., et al. 2012, MNRAS, 422, 3460
  \bibitem{} Schrijvers, C., Telting, J. H., De Ridder, J., 2002, ASP Conf. Ser., 259, 204
  \bibitem{} Stankov, A., Handler, G., 2005, ApJSuppl., 158, 193
  \bibitem{} Szewczuk, W., Daszy\'nska-Daszkiewicz, J., 2016, submitted to MNRAS
  \bibitem{} Waelkens, C., 1991, A\&A, 246, 453
  \bibitem{} Walczak, P., Fontes, Ch. J., Colgan, J., Kilcrease, D. P., Guzik, J. A., 2015, A\&A, 580, L9
\end{thebibliography}

\end{document}